\documentclass[prd,amsfonts,twocolumn,nofootinbib,showpacs]{revtex4}

\usepackage{graphicx}
\usepackage{dcolumn}
\usepackage{bm}

\def\sigv{\langle\sigma v\rangle}

\pacs {26.35.+c, 98.80.Cq, 98.80.Ft \hfill SACLAY--T09/046}

\begin{document}

\title{CMB constraints on Dark Matter models with large annihilation cross-section}

\author{Silvia Galli$^{a,b}$, Fabio Iocco$^{c,d}$, Gianfranco Bertone$^d$, Alessandro Melchiorri$^b$}
\affiliation{$^a$Laboratoire Univers et Th\'eorie (LUTH), Observatoire de Paris, Meudon, Universit\'e Paris Diderot - 75205 PARIS}
\affiliation{$^b$ Physics Department and INFN, Universita' di Roma ``La Sapienza'', Ple Aldo Moro 2, 00185, Rome, Italy}
\affiliation{$^c$ Institut de Physique Th\'eorique, CNRS, URA 2306 \& CEA/Saclay, F-91191 Gif-sur-Yvette}
\affiliation{$^d$ Institut d`Astrophysique de Paris, UMR 7095-CNRS Paris, Universit\'e Pierre et Marie Curie, boulevard Arago 98bis, 75014, Paris, France.}

\date{April 30, 2009}

\begin{abstract}
The injection of secondary particles produced by Dark Matter (DM)
annihilation around redshift $\sim 1000$ would inevitably
affect the process of recombination, leaving
an imprint on Cosmic Microwave Background (CMB) anisotropies and
polarization. We show that the most recent CMB measurements provided
by the WMAP satellite mission place interesting constraints on DM
self-annihilation rates, especially for models that exhibit a large Sommerfeld
enhancement of the annihilation cross-section, as recently
proposed to fit the PAMELA and ATIC results. Furthermore, we
argue that upcoming CMB experiments
such as Planck, will improve the constraints by at least one
order of magnitude, thus providing a sensitive probe of the
properties of DM particles.
\end{abstract}


\maketitle
\section{Introduction}
\label{intro}

The recent measurements of the Cosmic Microwave Background (CMB)
anisotropy and polarization from experiments as WMAP \cite{wmap5komatsu},
ACBAR \cite{acbar} and BOOMERANG \cite{boom} have confirmed the
theoretical predictions  of the standard cosmological model
based on inflation, dark matter and a cosmological constant.
This not only permits to place stringent constraints
on several parameters of the model but also to use it as new laboratory
where to test physical processes in a environment not 
achievable otherwise.

In particular, there is a remarkable agreement between the
theoretical description of the recombination process, occurring
at $z_r \sim 1000$, and CMB data, a circumstance that severely constrains new sources of ionizing
photons, and more in general any deviation from standard
recombination \cite{peeblesmod}, as recently shown by several
groups of authors (see e.g. \cite{galli}, \cite{naselski}, \cite{lewisweller}).
Most of the recent literature
has analyzed the modified recombination by means of a phenomenological approach,
parameterizing in a model independent way the modifications to the recombination process.
Here, we focus instead on the constraints that can
be placed on the properties of DM particles, under the assumption that
recombination is modified {\it only} by dark matter annihilation.
With respect previous studies \cite{dmannmodrec, base}, our
analysis includes more recent data
(WMAP 5-year data), and it concentrates on
a new class of DM models that have been recently proposed
to explain the observed anomalies in cosmic ray data.

In fact, the attempt to explain the high energy positron and
electron rise seen by PAMELA \cite{Adriani:2008zr} and ATIC \cite{ATIC:2008zzr}
in terms of Dark Matter (DM) annihilation
has prompted the proliferation of new DM candidates with very large
annihilation cross-section. In particular, in models with
a ``Sommerfeld'' enhancement of the annihilation cross-section $(\sigma v)$,
the efficent exchange of force carriers at low relative particle velocities
leads to a velocity-dependent $(\sigma v)$, which behaves roughly as
$ \propto 1/v$ for high $v$, and saturates below a critical
$v_s$ (typically smaller than the local velocity dispersion,
$v_\odot$, see below), that depends on the ratio between the masses
of the force carrier and the DM particle. A nice feature of these models is
that they can be made
naturally consistent with standard thermal freeze-out. In fact,
DM freezes out typically with $\beta \equiv v/c = \cal O$(1), and $(\sigma v)$ will grow from
this minimum value as the universe cools and expand.
Then, when the first gravitationally bound structures form,
DM virializes within the gravitational potential of the host halo,
thus leading for Milky Way (MW)-like galaxies at z=0 to virialized velocities
of order $\beta \sim 10^{-3}$. Smaller velocities, thus a larger
$(\sigma v)$ can be achieved in DM haloes with low velocities such as 
MW subhaloes\cite{SommDMsubMW} or small haloes at high redshift\cite{Kamionkowski:2008gj}.
although it is unclear whether the
annihilation flux can be boosted enough to explain the PAMELA
and ATIC data without being in conflict with other measurements,
such as the anti-proton or gamma-ray fluxes towards the
Galactic center (see e.g. \cite{bert1} and
references therein). 

When recombination occurs, around $z_r\sim$1000, the relic WIMPs have not
yet formed sizable gravitationally bound structures and are cold enough
for the Sommerfeld mechanism to produce substantial enhancement
of the annihilation cross-section with respect to the
thermal value (after kinetic decoupling
DM particle temperature evolves adiabatically
as T$\propto z^2$, so $\beta(z_r) \sim 10^{-8}$,
for a $\cal O$(100GeV/c$^2$) mass WIMP).
As we will see below, the actual enhancement is
model-dependent, because different DM models lead to a different
behaviour of $(\sigma v)_z$, but in general we expect that for
large enough cross-sections, DM annihilation will significantly
modify the recombination history, thus leaving a clear imprint on
the angular power spectra of CMB anisotropy and polarization.

Our paper is organized as follows: in the next section we describe
the effects of annihilating dark matter on primordial recombination and
the characteristic imprints on the CMB angular spectra. In section III
we describe our analysis method. In section IV we present the results
of our analysis. Finally, in Section V, we discuss our conclusions.

\section{Annihilating Dark Matter and Thermal History of the Universe}
\label{ionhist}

Annihilating particles affect the ionization hystory of the Universe in three main
different ways.
The interaction of the shower produced by the annihilation with the thermal gas can
{\it i}: ionize it, {\it ii}: induce Ly--$\alpha$ excitation of the hydrogen and {\it iii}: heat
the plasma; the first two  modify the evolution of the free electron fraction $x_e$,
the third affects the temperature of baryons.
In the presence of annihilating particles, the evolution of the
ionization fraction $x_e$ satisfies:
\begin{equation}
\label{eq:dxe}
\frac{d x_e}{d z} = \frac{1}{(1+z)H(z)}
\left[ R_s(z) - I_s(z) - I_X(z)  \right],
\end{equation}
where $R_s$ is the standard recombination rate, $I_s$ the ionization
rate by standard sources, and $I_X$ the ionization rate due to
 particle annihilation.

Following the seminal papers \cite{recbase}, standard recombination is described by:

\begin{equation}
   \left[ R_s(z) - I_s(z) \right]=C \times{\big[x_{\rm e}^2 n_{\rm H} \alpha_{\rm B}
 - \beta_{\rm B} (1-x_{\rm e})
   {\rm e}^{-h_{\rm P}\nu_{2s}/k_{\rm B}T_{\rm b}}\big]}
\end{equation}

where $n_H$ is the number density of hydrogen nuclei, $\alpha_{\rm B}$ and $\beta_{\rm B}$ are
the effective recombination and photo-ionization rates for principle
quantum numbers $\ge 2$ in Case B recombination, $\nu_{2s}$ is the frequency of the $2s$ level from
the ground state and $T_b$ is the temperature of the baryonic gas
and the factor $C$ is given by:

\begin{eqnarray}
C= \frac{\big[1 + K \Lambda_{2s1s} n_{\rm H}(1-x_{\rm e})\big]}
 {\big[1+K \Lambda_{2s1s} n_{\rm H} (1-x_{\rm e})
 + K \beta_{\rm B} n_{\rm H}(1-x_{\rm e})\big]}
\label{eq:standard_xe}
\end{eqnarray}
where   $\Lambda_{1s2s}$ is
the decay rate of the metastable $2s$ level, $n_{\rm H}(1-x_{e})$ is
the number of neutral ground state $H$ atoms, and $K=\lambda_\alpha^{3}/ (8\pi H(z))$
with $H(z)$ the Hubble expansion factor at redshift $z$ and $\lambda_{\alpha}$ is the wavelength of the Ly-$\alpha$
transition from the $2p$ level to the $1s$ level.

The $I_X$ term  of equation \ref{eq:dxe}
represents the contribute to the electron fraction evolution
by a ``non--standard'' source; in our case
it takes into account that during recombination particle
annihilation increases the ionization rate both by direct ionization
from the ground state, and by contributing additional Lyman-$\alpha$ photons.
The latter boosts the population at $n= 2$, increasing the rate of
photoionization by the CMB from these excited states. Therefore, the ionization rate due to particle annihilation is:
\begin{equation}
I_X(z) = I_{Xi}(z) + I_{X\alpha}(z) ,
\label{terms}
\end{equation}
where $I_{Xi}$ is the ionization rate due to ionizing photons, and $I_{X\alpha}$
the ionization rate due to additional Lyman alpha photons.

The rate of energy release $\frac{dE}{dt}$ per unit volume
by a relic self-annihilating dark matter particle is given by

\begin{equation}
\label{enrateselfDM}
\frac{dE}{dt}(z)= \rho^2_c c^2 \Omega^2_{DM} (1+z)^6 f  \frac{\sigv}{m_\chi}
\end{equation}

with $n_{DM}(z)$ being the relic DM abundance at a given redshift $z$,
$\sigv$ is the effective self-annihilation rate and $m_\chi$ the mass
of our dark matter particle, $\Omega_{DM}$ is the dark matter density
parameter and $\rho_c$ the critical density of the Universe today;
the parameter $f$ indicates the fraction of energy which is absorbed
{\it overall} by the gas, under the approximation the energy absorption
takes place locally. This on--the--spot approximation has been
adopted by previous analysis (\cite{base})

Each of the terms in Eq. \ref{terms} is related to the rate of energy release as:

\begin{eqnarray}
I_{Xi}&=&\phantom{(1-}\;\,C\phantom{)}\; \chi_i\frac{[dE/dt]}{n_H(z) E_i} \\
I_{X\alpha} &=&(1-C)\; \chi_\alpha\frac{[dE/dt]}{n_H(z) E_\alpha} \\
\end{eqnarray}
 where $E_i$ is the average ionization energy per baryon, $E_\alpha$ is the difference
in binding energy between the $1s$ and $2p$
energy levels of a hydrogen atom, $n_H$ is the number density of Hydrogen Nuclei and
$\chi_i=\chi_\alpha=(1-x_e)/3$ are the fractions of energy going to ionization and to
Lyman-$alpha$ photons respectively, given by \cite{CK04} following the approach of Shull and Van Steenberg \cite{ShullVanSteen1985}.

Finally, a fraction of the energy released by annihilating particles goes into heating of baryonic gas, adding an extra $K_h$ term in the standard evolution equation for the matter temperature $T_b$:

\begin{eqnarray}
(1+z)\frac{dT_b}{dz}&=&\frac{8\sigma_T a_R T_{CMB}^4}{3m_e
c H(z)}\frac{x_e}{1+f_{\rm He}+x_e} (T_b -T_{CMB})\nonumber \\
&& -\frac{2}{3
k_B H(z)} \frac{K_h}{1+f_{\rm He}+x_e} +2 T_b,
\end{eqnarray}
where the non standard term is given by:
\begin{equation}
K_h=\chi_h \frac{(dE/dt)}{n_H(z)}
\end{equation}
and $\chi_h =(1+2x_e)/3$ is the fraction of energy going into heat given by \cite{CK04}.

\section{Annihilating Dark Matter and the CMB}
\label{ionhistcmb}

\begin{figure}[t]
\centering
 \includegraphics[angle=0,width=0.4\textwidth]{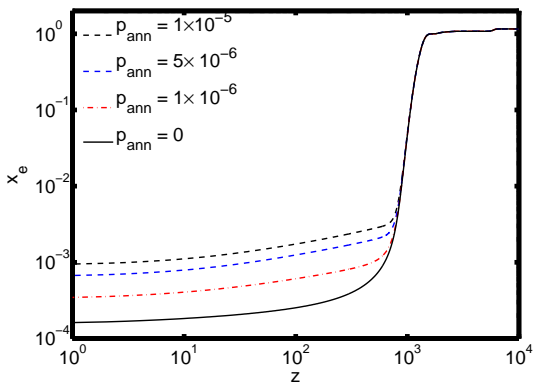}
\caption{Evolution of the free electron fraction as function of redshift for different
values of $p_{ann}=[0, 10^{-6},5\times10^{-6},10^{-5}]$ $m^3/s/Kg$.}
\label{xe}
\end{figure}

\begin{figure}[t]
\centering
 \includegraphics[width=0.4\textwidth]{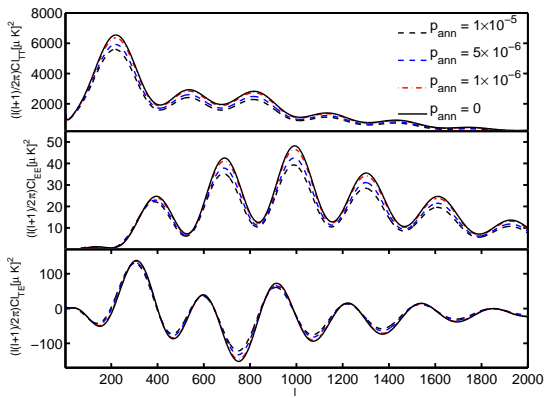}
\caption{TT, TE, EE angular power spectra (from Top to Bottom) for different
values of $p_{ann}=[0, 10^{-6},5\times10^{-6},10^{-5}]$ $m^3/s/Kg$ . }
\label{TTEE}
\end{figure}

We can now compute the theoretical angular power in presence of DM annihilations,
by modifying the RECFAST routine (\cite{recfast})
in the CAMB code (\cite{camb}), following the prescription described in the previous section. The
dependence on the properties of the DM particles is encoded in the quantity
\begin{equation}
f\frac{<\sigma v>}{m_\chi} \equiv p_{ann}
\label{pann}
\end{equation}
appearing in eq. \ref{enrateselfDM}, that we use as a parameter in the code.

In Fig \ref{xe} we show the evolution of the free electron fraction for
different values of ${p_{ann}}$. As we can see, the DM annihilation model we
consider can increase the free electron fraction after $z\sim 1000$ by one order
of magnitude, increasing the optical depth to last scattering surface and
smearing the visibility function.
The consequences of such annihilation can be seen in Fig.\ref{TTEE}
where we show the CMB anisotropy, cross-polarization
and polarization angular power spectra for different values of  ${p_{ann}}$.
DM annihilation damps the acoustic oscillations in the angular power spectra
as in the case of an instantaneous reionization. However, large scale polarization is
left unchanged by dark matter annihilation and a degeneracy between
these two effects can indeed be broken. Although DM annihilation could play a 
role in the subsequent reionization of the Universe , the effect
is likely to be small \cite{reionDM}, unless one invokes very high anihilation cross sections \cite{reionDM2}. 
Here, we don't consider a particular model for reionization, and simply adopt the 
parametrization of a full and instantaneous reionization at redshift $z_r<30$.

\begin{figure}[t]
\centering
 \includegraphics[width=0.4\textwidth]{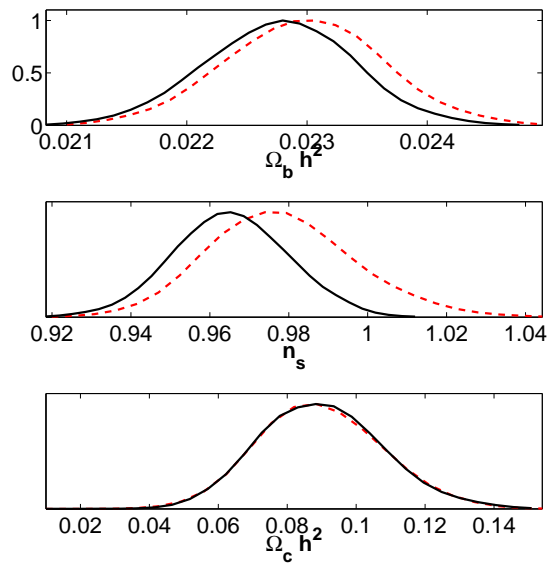}
\caption{Constraints on the $\omega_b$, $n_s$ and $\omega_c$ parameters
in the case of standard recombination (solid line), or including dark matter annihilation
(dashed line).}
\label{obns}
\end{figure}

\begin{table}[b]
\begin{center}
\begin{tabular}{rc}
Experiment & $p_{ann}$ $95 \%$ c.l.\\
\hline
WMAP& $< 2.0\times10^{-6}$m$^3$/s/kg\\
\hline
Planck& $< 1.5\times10^{-7}$m$^3$/s/kg\\
\hline
CVl& $<  5.0\times10^{-8}$ m$^3$/s/kg\\
\hline
\end{tabular}
\caption{Upper limit on $p_{ann}$ from current WMAP observations and future upper limits achievable
from the Planck satellite mission and from a cosmic variance limited experiment.}
\label{tab:exp}
\end{center}
\end{table}

We search for an imprint of self-annihilating dark matter in current CMB angular spectra
by making use of the publicly available Markov Chain Monte Carlo
package \texttt{cosmomc} \cite{Lewis:2002ah}.
Other than $p_{ann}$ we sample the following six-dimensional set of cosmological parameters,
adopting flat priors on them:
the physical baryon and CDM densities, $\omega_b=\Omega_bh^2$ and
$\omega_c=\Omega_ch^2$, the scalar
spectral index, $n_{s}$,
the normalization, $\ln10^{10}A_s(k=0.05/Mpc)$,
the optical depth to reionization, $\tau$, and the
ratio of the sound horizon to the angular diameter distance,
$\theta$.

We consider purely adiabatic initial conditions.
The MCMC convergence diagnostic tests are performed on $4$ chains using the
Gelman and Rubin ``variance of chain mean''$/$``mean of chain variances'' $R-1$
statistic for each parameter. Our $1-D$ and $2-D$ constraints are obtained
after marginalization over the remaining ``nuisance'' parameters, again using
the programs included in the \texttt{cosmomc} package.
We use a cosmic age top-hat prior as 10 Gyr $ \le t_0 \le$ 20 Gyr.
We include the five-year WMAP data \cite{wmap5komatsu} (temperature
and polarization) with the routine for computing the likelihood
supplied by the WMAP team (we will refer to this analysis as WMAP5).

\section{Results}

\begin{figure}[t]
\centering
 \includegraphics[width=0.5\textwidth]{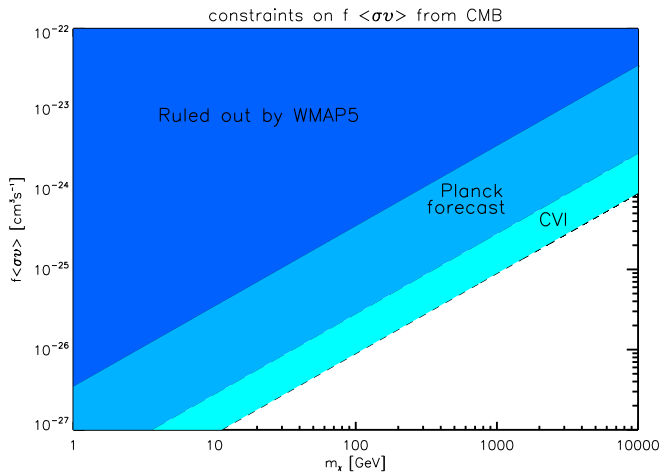}
\caption{Constraints on the self-annihilation cross-section at recombination $(\sigma v)_{z_r}$
times the gas--shower coupling parameter $f$. The dark blue area is already excluded by WMAP5 data,
whereas the more stringent limit (dashed area) refers to the constraints which will be possible to apply with Planck.
The light blue area is the zone ultimately allowed to probe by a cosmic variance limited experiment
with angular resolution comparable to Planck.}
\label{fsigmav}
\end{figure}

Using the WMAP-5 dataset and applying the analysis method described
in the previous section, we found an upper limit
$p_{ann}<2.0\times10^{-6}$ m$^3$/s/kg at $95 \%$ c.l., with no indications
for modified recombination in agreement with previous and
similar analyses. The implications of this limit are discussed in the next section.
While we detect only an upper limit it is interesting, from a cosmological point
of view, to investigate the possible impact of this parameter on the
estimation of other parameters as the baryon density $\omega_b$,
the cold dark matter density $\omega_c$  and
the scalar spectral index $n_S$. In Figure \ref{obns} we plot the $1$-D likelihood
distributions for these three parameters derived assuming the standard
case (i.e. $p_{ann}=0$) and letting this parameter to vary freely.
As we can see, including $p_{ann}$ into the analysis changes
the constraints of $\omega_b=0.0228\pm0.0006$ and $n_s=0.965\pm0.014$ (obtained
in the standard case) to $\omega_b=0.0230\pm0.0006$ and $n_s=0.977\pm0.018$.
The cosmological constraints on the cold dark matter density are on
the contrary not affected by the inclusion of $p_{ann}$.

With the advent of the Planck satellite mission, it is interesting to
forecast to what extent the above limit will be improved by this
mission. We have therefore forecasted future constraints on $p_{ann}$ assuming
simulated Planck mock data with a fiducial model
given by the best fit WMAP5 model (with standard recombination)
and experimental noise described by (see \cite{planck}):
\begin{equation}
N_{\ell} = \left(\frac{w^{-1/2}}{\mu{\rm K\mbox{-}rad}}\right)^2
\exp\left[\frac{\ell(\ell+1)(\theta_{\rm FWHM}/{\rm rad})^2}{8\ln 2}\right],
\end{equation}
with $w^{-1/2}=63 \mu K$ as the temperature noise level
(we consider a factor $\sqrt{2}$ larger for polarization
noise) and  $\theta_{\rm FWHM}=7'$ for the beam size.
We take $f_{sky}=0.65$ as sky coverage.
We found that the Planck mission in the configuration
described above will have the ability of placing a constraint of
$p_{ann}<1.5\times10^{-7}$ m$^3$/s/kg at $95 \%$ c.l.

It is also interesting to investigate the ultimate ability
of cosmology to place constraints on $p_{ann}$. We have therefore
repeated the analysis with an ideal Cosmic Variance Limited experiment
with resolution up to $\ell_{max}=2500$. In this case we found
$p_{ann}< 5.0\times10^{-8}$ m$^3$/s/kg at $95 \%$ c.l.

These constraints are summarized in fig. \ref{fsigmav}, where we show
the allowed values of $f\sigv$ as a function of the WIMP mass $m_\chi$,
for the different experiments described above.
These results place useful constraints on the DM annihilation
cross-section at very small relative velocity. This is particularly
important for models with a large ``Sommerfeld enhancement" (SE), a
non-perturbative effect arising from the distortion of the
wave functions of the two annihilating particles, due to the exchange
of Coulomb-like forces mediated by (possibly new) force carriers
\cite{Somm}. The interest in these models arises from the
fact that larger-than-thermal annihilation
cross-section are required if one wants to explain the
the rise in the electron and positron spectra observed
by PAMELA and ATIC in terms of DM annihilation
(see e.g. the discussion in Ref.~\cite{Cirelli:2008pk}).
We briefly recall here the basics of the SE.
For two DM particles undergoing s-wave annihilation, the
wave function in the non-relativistic limit obeys the
Schr\"odinger equation
\begin{equation}
\psi''(r) - m_\chi V(r)\psi(r) + m_\chi^2 \beta^2 \psi(r) =0
\label{schroe}
\end{equation}
In the limit where the mass of the carrier
and the relative velocity of DM particles are small, it is
easy to find an analytic approximation to the SE
\begin{equation}
S(\beta)=\frac{\alpha \pi}{\beta} [1-\exp^{-\alpha \pi/\beta}]
\end{equation}
which exhibits the $S \sim 1/\beta$ behaviour that we mentioned
in the introduction.
\begin{figure*}[!t]
\centering
 \includegraphics[width=0.8\textwidth]{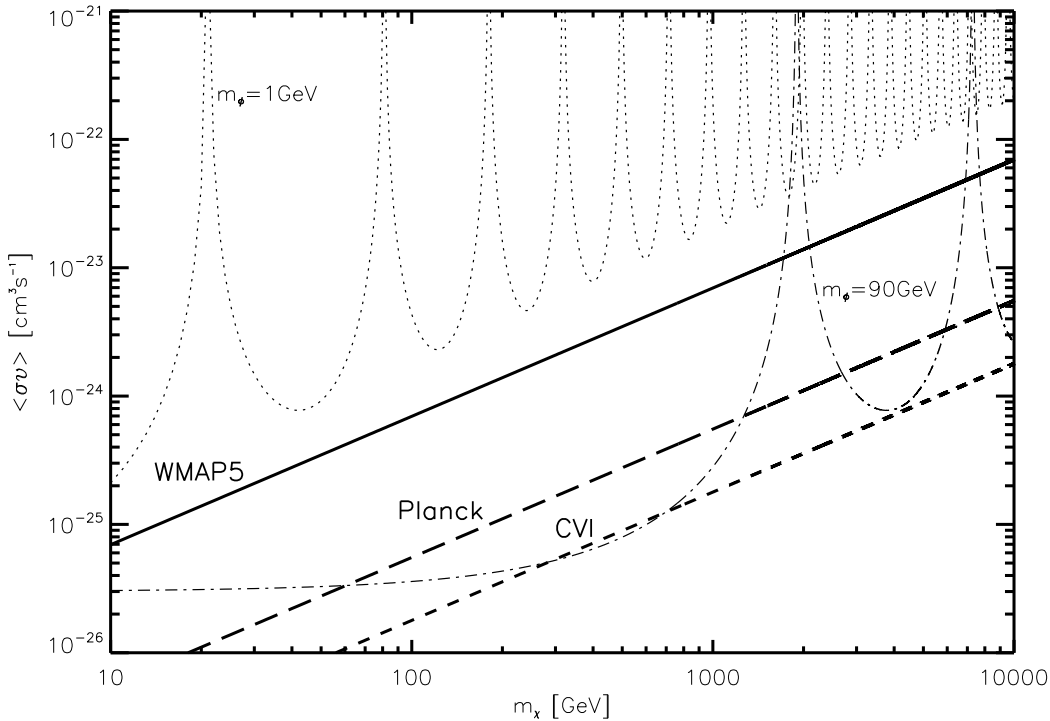}
\caption{Constraints on the self-annihilation cross-section at
 recombination $(\sigma v)_{z_r}$, assuming the gas--shower coupling parameter $f$=0.5, see text for details.
Regions above the solid (/long dashed/short dashed) thick lines are ruled out by WMAP5
(/Planck forecast/Cosmic Variance limited);
the thin dotted and dashed-dotted lines are the predictions of the ``Sommerfeld''
enhanced self--annihilation cross sections with force carrying
bosons of m$_\phi$=1GeV/c$^2$ and m$_\phi$=90GeV/c$^2$ respectively,
see text for details. Notice that these constraints apply to $\sigv$ at very low
temperatures such that it is in saturated  Sommerfeld regime, and therefore
directly comparable with results from galatic  substructures and dwarf galaxies
constraints as from \cite{SommDMsubMW}. }
\label{sigmav}
\end{figure*}
Interestingly, a full calculation shows that the true
solution saturates at $\beta \sim m_\phi/m_\chi$,
and it actually develops resonances, that lead to very
large SE for specific combinations of masses $m_\phi$ and
$m_\chi$, and the coupling $\alpha$.
In order to compare the constraints on $p_{ann}$ obtained from the
analysis of CMB data with theoretical models, we have
numerically integrated equation \ref{schroe}, assuming
a Yukawa potential $V(r)=-\exp[-m_\phi r]\alpha/r$
mediated by a boson of mass $m_\phi = 1$GeV/c$^2$, and
$m_\phi = 90$GeV/c$^2$, taking $\alpha$=1/4$\pi$
(see e.g. \cite{ArkaniHamed:2008qn} for details);
$\beta$=10$^{-8}$ as appropriate for the recombination epoch.

The results are visualized in fig.\ref{sigmav}, were we show
the region excluded by our analysis
in the $(\sigma v)$ vs. $m_\chi$ plane, corresponding to
the 95 \% c.l. upper limit on the cross section that can be derived 
by combining eq. \ref{pann} with the constrains on $p_{ann}$ in
table \ref{tab:exp}:
\begin{equation}
\sigma v_{z_r,26}^{\rm{max}} = 71.2 \left( \frac{p_{ann}^{\rm{max}}}{2.0\cdot 10^{-6} \rm{m}^3\rm{s}^{-1}\rm{kg}^{-1}} \right) \left( \frac{m_\chi}{100 \rm{GeV}} \right) \left( \frac{0.5}{f} \right)
\end{equation}
where $\sigma v_{z_r,26}^{\rm{max}}$ denotes the upper limit of
the annihilation cross section at recombination in units of 
$10^{-26}$ cm$^3$s$^{-1}$. 

We have adopted in this formula, and in fig.\ref{sigmav}, a fiducial value 
$f=0.5$ for the coupling between the annihilation
products and the gas,
following the detailed calculation of DM--induced shower
propagation and energy release performed by \cite{FinkInProg};
this number is a good approximation averaged on all channels, although
its actual value will ultimately depend on the composition of the
shower and on its energy spectrum, namely on the nature
of the annihilating DM particle itself.
It is however straightforward to obtain
the constraints for different values of $f$.

We find that the most extreme enhancements are already
ruled out by existing CMB data, while enhancements of
order $10^3$--$10^4$ with respect to thermal
value $\sigv$=3$\times$10$^{-26}$ cm$^3$/s, required to explain the
PAMELA and ATIC data, will be probed over a larger
WIMP mass range by Planck.
We also note that  for small enough $m_\chi$, a CMB experiment allows
us to probe the region of thermal cross-sections, and that Planck
sensitivity will reach it, making it possible perhaps to find hints
of particle DM in CMB data.

We note that the constraints obtained here are several orders of magnitude
more stringent than those obtained from the analysis of high-redshift 
proto-halos~\cite{Kamionkowski:2008gj}(see also the recent \cite{CyrRacine:2009yn}).
We stress that our results apply also for standard models with s-wave 
annihilations, where the annihilation cross section does not depend on $v$.
In this case, our results can be directly compared with the constraints
from astrophysical observations in the local universe~\cite{Bell2008}.

\section{Conclusions}

We have studied the effects of WIMP DM self-annihilation on
recombination, looking for signatures in the CMB anisotropy
and polarization. Our analysis has been performed under the
assumption that the shower produced by the WIMP annihilation
interacts ``locally'' with plasma, and a fraction $f$ of the energy
is absorbed on--the--spot by the baryons, contributing to
its ionization. Our methodology is consistent with other analysis
on the effect of decaying or  self--annihilating, low--mass dark
matter performed in the past.
We examine a range of higher WIMP masses (1 GeV/c$^2$--10 TeV/c$^2$),
and find that current WMAP data already allow us to put
interesting constraints on self--annihilation cross sections
higher than the ``standard'' thermal value, in the range of those
required to explain the PAMELA and ATIC data in terms of
dark matter.  Physically motivated by the very low relative velocity
of DM particles at the time of recombination, our constraints
on ``Sommerfeld'' enhanced cross sections are competitive with
the existing ones from local Universe (galactic substructures),
and an independent test achieved with standard physics of the
early Universe.
By using simulated mock data, we have found that the expected enhanced
sensitivity of the Planck mission will improve our capability to
constrain Sommerfeld enhancement in dark matter particle
models, thus hinting toward the exciting possibility
to be finding traces of particle dark matter in future
CMB data.
Interestingly, Planck will very likely be able to probe the region
of the thermal annihilation cross--section for low WIMP
masses ($\lesssim$50GeV/$c^2$), the actual value depending on the
gas--shower coupling $f$.
Ultimately, a cosmic variance limited experiment
 permits the possibility to probe cross-sections values at
the order of the thermal one for ${\cal O}$(100 GeV/c$^2$)
mass WIMPs.

\acknowledgments
We are glad to acknowledge fruitful conversations with Marco Cirelli,
Douglas Finkbeiner and Pasquale D.~Serpico. S.~G. and A.~M. would
like to thank M. Kamionkowski for useful comments.
This research has been partially supported by ASI contract I/016/07/0 ``COFIS.''

\newcommand\AAP[3]{Astron. Astrophys.{\bf ~#1}, #2~ (#3)}
\newcommand\AL[3]{Astron. Lett.{\bf ~#1}, #2~ (#3)}
\newcommand\AP[3]{Astropart. Phys.{\bf ~#1}, #2~ (#3)}
\newcommand\AJ[3]{Astron. J.{\bf ~#1}, #2~(#3)}
\newcommand\APJ[3]{Astrophys. J.{\bf ~#1}, #2~ (#3)}
\newcommand\ApJ[3]{Astrophys. J.{\bf ~#1}, #2~ (#3)}
\newcommand\APJL[3]{Astrophys. J. Lett. {\bf ~#1}, L#2~(#3)}
\newcommand\APJS[3]{Astrophys. J. Suppl. Ser.{\bf ~#1}, #2~(#3)}
\newcommand\MNRAS[3]{Mon. Not. R. Astron. Soc.{\bf ~#1}, #2~(#3)}
\newcommand\MNRASL[3]{Mon. Not. R. Astron. Soc.{\bf ~#1}, L#2~(#3)}
\newcommand\NPB[3]{Nucl. Phys. B{\bf ~#1}, #2~(#3)}
\newcommand\PLB[3]{Phys. Lett. B{\bf ~#1}, #2~(#3)}
\newcommand\PRL[3]{Phys. Rev. Lett.{\bf ~#1}, #2~(#3)}
\newcommand\PR[3]{Phys. Rep.{\bf ~#1}, #2~(#3)}
\newcommand\PRD[3]{Phys. Rev. D{\bf ~#1}, #2~(#3)}
\newcommand\SJNP[3]{Sov. J. Nucl. Phys.{\bf ~#1}, #2~(#3)}
\newcommand\ZPC[3]{Z. Phys. C{\bf ~#1}, #2~(#3)}
\newcommand\SCI[3]{Sci.{\bf ~#1}, #2~(#3)}

\end{document}